\documentclass[twocolumn]{aastex61}

\usepackage{graphicx}
\usepackage{natbib}

\submitjournal{ApJ}

\begin{document}

\title{GASP IV: A muse view of extreme ram-pressure stripping in the plane of the sky: the case of jellyfish galaxy JO204}

\shorttitle{GASP IV: JO204}
\shortauthors{Gullieuszik et al.}

\correspondingauthor{Marco Gullieuszik}
\email{marco.gullieuszik@oapd.inaf.it}

\author{Marco Gullieuszik}
\affiliation{INAF--Astronomical Observatory of Padova, vicolo dell'Osservatorio 5, I-35122 Padova, Italy}

\author{Bianca M. Poggianti}
\affiliation{INAF--Astronomical Observatory of Padova, vicolo dell'Osservatorio 5, I-35122 Padova, Italy}

\author{Alessia Moretti}
\affiliation{INAF--Astronomical Observatory of Padova, vicolo dell'Osservatorio 5, I-35122 Padova, Italy}

\author{Jacopo Fritz}
\affiliation{Instituto de Radioastronom\'ia y Astrof\'isica, IRyA, UNAM, Campus Morelia, A.P. 3-72, C.P. 58089, Mexico}
  
\author{Yara L. Jaff\'e}
\affiliation{European Southern Observatory, Alonso de C\'ordova 3107, Santiago, Chile}

\author{George Hau}
\affiliation{European Southern Observatory, Alonso de C\'ordova 3107, Santiago, Chile}

\author{Jan C. Bischko}
\affiliation{Institute of Astro- and Particle Physics, University of Innsbruck, Technikerstrasse 25, 6020 Innsbruck, Austria}

\author{Callum Bellhouse}
\affiliation{European Southern Observatory, Alonso de C\'ordova 3107, Santiago, Chile}

\author{Daniela Bettoni}
\affiliation{INAF--Astronomical Observatory of Padova, vicolo dell'Osservatorio 5, I-35122 Padova, Italy}

\author{Giovanni Fasano}
\affiliation{INAF--Astronomical Observatory of Padova, vicolo dell'Osservatorio 5, I-35122 Padova, Italy}

\author{Benedetta Vulcani}
\affiliation{School of Physics, University of Melbourne, VIC 3010, Australia}
\affiliation{INAF--Astronomical Observatory of Padova, vicolo dell'Osservatorio 5, I-35122 Padova, Italy}

\author{Mauro D'Onofrio}
\affiliation{Department of Physics and Astronomy, University of Padova, Vicolo Osservatorio 3, I-35122 Padova, Italy}

\author{Andrea Biviano}
\affiliation{INAF--Astronomical Observatory of
Trieste, via G.B. Tiepolo 11, 34131 Trieste, Italy}

\begin{abstract}

In the context of the GAs Stripping Phenomena in galaxies with Muse (GASP) survey, we present the characterization of JO204, a jellyfish galaxy in A957, a relatively low-mass cluster with $M=4.4 \times10^{14}M_\sun$. This galaxy shows a tail of ionized gas that extends up to 30 kpc from the main body in the opposite direction of the cluster
center. No gas emission is detected in the galaxy outer disk,
suggesting that gas stripping is proceeding outside-in. The stellar
component is distributed as a regular disk galaxy; the  stellar kinematics shows a symmetric rotation curve with a maximum radial velocity of 200km/s out to 20 kpc from the galaxy center. The radial velocity of the gas component in the central part of the disk follows the distribution of the stellar component; the gas kinematics in the tail retains the rotation of the galaxy disk, indicating that JO204 is moving at high speed in the intracluster medium. Both the emission and radial velocity maps of the gas and stellar components indicate
ram-pressure as the most likely primary mechanism for gas stripping,
as expected given that JO204 is close to the cluster center and it is
likely at the first infall in the cluster.
The spatially resolved star formation history of JO204 provides evidence that
the onset of ram-pressure stripping occurred in the last 500 Myr, quenching the star formation activity in 
the outer disk, where the gas has been already completely stripped.
Our conclusions are supported by a set of hydrodynamic simulations.

\end{abstract}


\section{Introduction} \label{sec:intro}
The evolution of galaxies is primarily regulated by the processes of
gas inflow and outflow. A galaxy without a reservoir of gas will
soon stop its star formation activity and will therefore evolve
into a red and passive galaxy.  \cite{peng+2010} showed that the
differential effects of galactic mass and of the environment in the
quenching of galaxies are fully separable, and introduced the idea of
two distinct processes operating: "mass quenching" and "environment
quenching".

The evolution of isolated galaxies is mainly driven by the balance of
gas consumption by star formation, gas loss by outflows and gas
accretion. The main driver for these processes is the galaxy mass; in
general, massive galaxies are more likely to consume all the gas
reservoir and become quiescent than low-mass galaxies.  The most
direct evidence of the environment effect in quenching the star
formation history is the morphology-density relation
\citep{dres+1980,fasa+2015}: the fraction of late-type galaxies
decreases from $\sim80\%$ in the field to $\sim60\%$ in the outskirts of
galaxy clusters to virtually zero in the cores of rich clusters.

Gravitational interactions 
are obvious candidate mechanisms that could efficiently remove the gas from galaxies;
these include strong tidal interactions and mergers between galaxies \citep{barn+1992},
tidal interactions with the cluster potential \citep{byrd+1990}, and the so-called 
harassment, which is the effect of multiple high-speed galaxy-galaxy close encounters \citep{moor+1996}.
Another class of possible quenching mechanisms is the
interactions of the cold disk gas with the hot
intracluster medium (ICM), like
thermal evaporation \citep{cowi+1977}, viscous stripping \citep{nuls+1982}
or ram-pressure stripping \citep{gunn+1972}.
Moreover, further infall of gas into the galaxy disk can be prevented by the removal of the outer galaxy gas halo (by either
hydrodynamic interaction with the ICM or tidal forces), thus causing the quenching by starvation\citep{lars+1980}.

In galaxy clusters, ram-pressure stripping is expected to be 
the most efficient of the aforementioned processes.
In the inner regions of rich clusters, galaxy-galaxy interactions
are expected to be inefficient and rare because of the large velocity
dispersion;
for the same reason, and because of the high density of the ICM,
this environment is instead considered highly favorable for
ram-pressure stripping \citep[e.g.][]{stei+2016}.
A clear observational signature of ongoing ram-pressure stripping is
an uneven gas distribution in the galaxy disk \citep{giov+1985,gava+1989}.
In the most extreme cases, extended one-sided gas tails of
gas can be observed to extend up to many disk scale lengths:
these spectacular examples of gas stripping galaxies are commonly referred
to as jellyfish galaxies \citep{cort+2007,smit+2010,ower+2012,ebel+2014}.
The first systematic searches of jellyfish galaxies were conducted
recently in nearby \citep{pogg+2016} and high-redshift ($z=0.3$ -- 0.7) clusters
\citep{ebel+2014,mcpa+2016}.

GASP\footnote{\url{http://web.oapd.inaf.it/gasp}} (GAs Stripping
Phenomena in galaxies with MUSE) is an ESO Large programme aimed at
studying where, how, and why gas can be removed from galaxies.  GASP
targets 94 candidate stripping galaxies selected from the
\citet{pogg+2016} catalog, which was built on a systematic search for
galaxies with signatures of  one-sided debris or tails reminiscent of
gas stripping processes in optical images of clusters from WINGS
\citep{fasa+2006} and OmegaWINGS \citep{gull+2015} surveys and groups
from the Padova Millennium Galaxy and Group Catalogue
\citep[PM2GC,][]{calv+2011}.
GASP was granted 120 hr of observing time with
the  integral-field spectrograph MUSE mounted at the VLT; observations
started in
October 2015 and --at the current rate of
execution-- we foresee the end of GASP observations by the end of 2018.
A complete description of the survey strategy, data reduction, and analysis procedures
is presented in \cite{pogg+2017}.

In this paper we present GASP observations of the galaxy named JO204
in \cite{pogg+2016}, which is J101346.82-005450.9 in the WINGS database
\citep{more+2014}. This galaxy turned out to be one of the most
spectacular cases of jellyfish galaxies observed by GASP to date.
The paper is organized as follows. In Sect. \ref{sec:obs} we present
GASP observations and data reduction;
in Sect. \ref{sec:ana} we 
describe the data analysis ; Sect. \ref{sec:env} describes JO204's environment;
in Sect. \ref{sec:sims} we present a set of hydrodynamic simulations of ram-pressure stripping
in a galaxy resembling JO204;
in Sect. \ref{sec:conclusions} we summarize the paper and provide our main conclusions.

In this paper we adopt a \cite{chab+2003} IMF and standard concordance cosmology with $H_0=70$
$\mathrm{km}\,\mathrm{s}^{-1}\,\mathrm{Mpc}^{-1}$, $\Omega_M=0.3$,
$\Omega_\Lambda=0.7$. This gives a scale of 0.887 kpc/arcsec at the
redshift of A957, that is $z=0.0451$ \citep{more+2017}.

\section{Observations and data reduction}\label{sec:obs}

All GASP observations are carried out in service mode with the MUSE
spectrograph \citep{baco+2010}, mounted at the Nasmyth focus of the UT4 VLT, at Cerro
Paranal in Chile.  MUSE is composed of 24 IFU modules and each of them
is equipped with a 4k$\times$4k CCD.  The spectral range, between
4500\AA\ and 9300\AA, is sampled at 1.25 \AA\//pixel with a spectral
resolution of $\sim2.6$\AA. The resolving power at 7000\AA\ is $R=2700$, corresponding
to 110 km/s or 53 km/s/pixel.
The 1\arcmin$\times$1\arcmin\ field of
view is sampled at $0\farcs2$/pixel; each datacube therefore consists
of $\sim10^5$ spectra.

\begin{figure}[t]
 \includegraphics[width=\hsize]{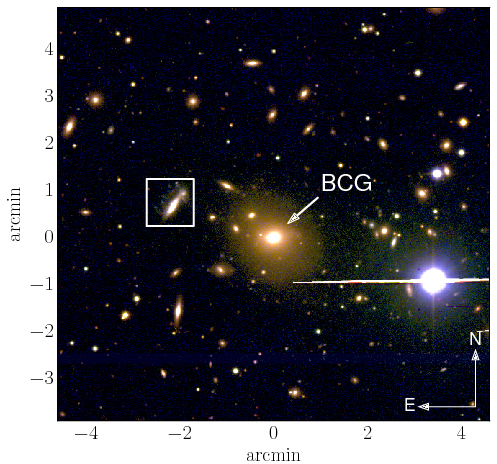}
  \caption{RGB image of the central region of A957 obtained from
    OmegaWINGS $u$- and WINGS $B$- and $V$-band images.  The footprint
    of the field observed by MUSE is shown by the white box.  The
    position of the brightest cluster galaxy is marked by the arrow.
\label{fig:a957}}
\end{figure}

JO204 is a member of the Abell 957 (A957) cluster and
it is located at $2\farcm1$ from the cluster center (see
Fig.\ref{fig:a957}), which corresponds to a projected distance of 132
kpc.  JO204 hosts a known AGN \citep[J101346.8-005451,][]{vero+2010}
with a double peaked [OIII] \citep{wang+2009}; this feature is
interpreted as either a binary AGN or a sign of an outflow \citep[see
  e.g.][]{mull+2015}.  JO204 was classified as a very probable
jellyfish galaxy (JClass=5) because in WINGS $B$-band images it shows
a unilaterally disturbed morphology, with filaments suggesting
ongoing ram-pressure gas stripping (see Fig. \ref{fig:rgb}).

For JO204 we adopted an observing strategy slightly different from the standard
GASP one \citep{pogg+2017}. Observations were split into three blocks, each consisting
of two 675 s exposures. The pointing of each block was shifted in the North-East direction
to extend the coverage of the JO204 gas tail (see Fig. \ref{fig:rgb}). 
Observations were carried out on three different nights --December 05 2015,
January 06 2016, and February 27 2016-- under good weather conditions
(the ESO DIMM recorded a seeing between $0\farcs7$ and $1\farcs0$).
Raw data were reduced using the
latest ESO MUSE pipeline available after the observations were
completed (v1.2.1)\footnote{The quality of the pipeline products
  fulfills  the scientific requirements of the GASP survey; thus we decided
  not to reprocess all raw data with a more recent version of the
  software.}, following the standard GASP data reduction procedure,
which is described in detail in \cite{pogg+2017}.

The final stacked MUSE datacube consists of $353\times369$
sky-subtracted, flux-calibrated spectra with radial velocities
corrected to the barycenter of the solar system.  The FWHM of
point-like sources on the image obtained by convolving the final MUSE
datacube with the $V$-band filter is $0\farcs8$ (4 pixels).

\section{Analysis}\label{sec:ana}

\begin{figure}[t]
  \includegraphics[width=\hsize]{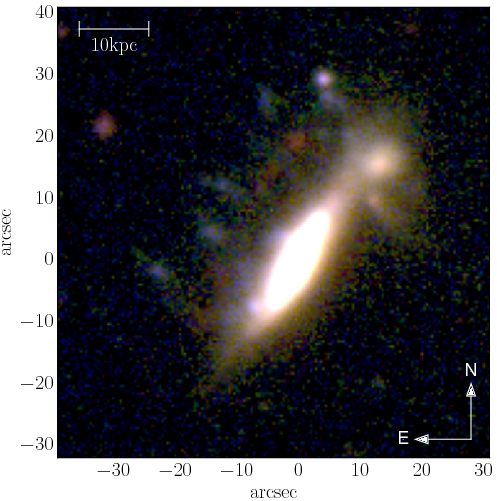}\\ \vfill
  \includegraphics[width=\hsize]{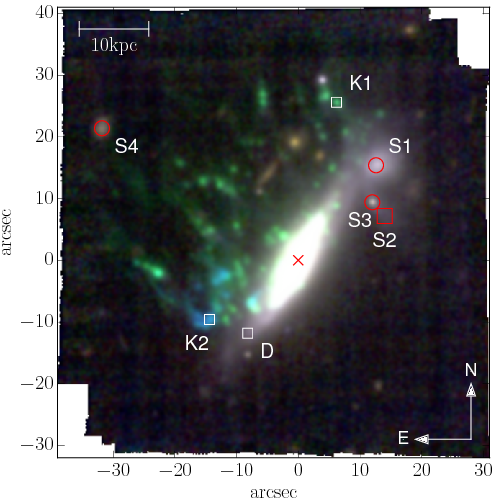}
  \caption{{\it Upper panel}: Zoom in of the image in
    Fig. \ref{fig:a957} on JO204.  {\it Lower panel:} RGB color image
    of JO204 obtained using [OIII]5007 for the blue channel,
    H$\alpha$ for the green one, and the continuum from 7100 and
    7200\AA\ for the red one. The
    sources marked by squares and circles are discussed in the
    text. The cross is the optical center, which was defined as the origin of
    the coordinate systems of all maps in this paper. 
    Areas in white are not covered by MUSE data; at the
    upper-right and lower-left corners one can see the
    effect of the offset applied to the pointing center of the three observing blocks.
\label{fig:rgb}}
\end{figure}

MUSE observations reveal that the faint extraplanar emission 
visible in optical broadband images (upper panel in Fig. \ref{fig:rgb}) is actually
an extended tail of ionized gas, with
strong $H\alpha$ emission.  This is evident from the RGB image
obtained from the MUSE datacube that is shown in the lower panel of
Fig. \ref{fig:rgb}.  The blue and the green channels were obtained by
integrating the MUSE datacube over a band of $\pm20$\AA\ centered on
the wavelength of [OIII]5007 and H$\alpha$ (rest-frame); for
the red channel we selected a band with no emission lines from JO204,
from 7100\AA\ to 7200\AA.  We selected the brightest pixel in this red image as
the center of JO204 and we defined it as the origin of
the spatial coordinate system for all maps presented in this paper.

JO204 shows extended regions of H$\alpha$ emission (green in
Fig. \ref{fig:rgb}), out to $\sim30$-40\arcsec\ ( $\sim30$kpc) from
the disk. In the filaments of diffuse H$\alpha$ emission there are
many compact emitting knots (an example is the one labeled K1 in
Fig. \ref{fig:rgb}), which are shown to be
active star-forming HII regions later in this paper.  In
the East part of the tail, approximately at
($-10$\arcsec,$-15$\arcsec) from the center of the galaxy, there is an
extended region with a prominent [OIII] emission, including an  extremely
bright compact region, labeled K2 in Fig. \ref{fig:rgb}.  The
differences in [OIII]/H$\alpha$ flux ratios across JO204 strongly
suggest that different ionizing mechanisms are at work.  The ionized
gas emission will be further discussed in Sect. \ref{sec:ion}. 

\begin{figure}
  \includegraphics[width=\hsize]{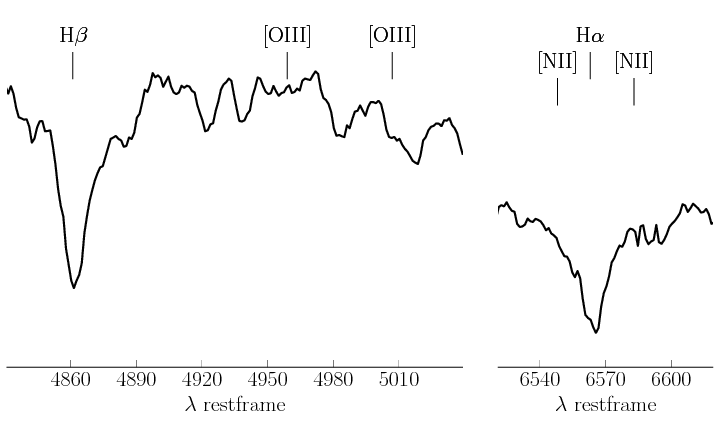}
  \caption{The k+a post-starburst spectrum (on an arbitrary scale)
    of a $1\arcsec$ region
    in the outer disk of JO204 (region D in Fig. \ref{fig:rgb}).
\label{fig:psb}}
\end{figure}

In the
outer South-East side of the disk --which is visible out to
$\sim$($-15$\arcsec,$-10$\arcsec)-- there is no gas emission, as demonstrated by the spectrum shown in Fig. \ref{fig:psb}, which was obtained by integrating the MUSE datacube over a $1\arcsec$ region at
$\sim -8^\prime, -12^\prime$ from the center (marked as D
in Fig. \ref{fig:rgb}).
This spectrum
shows a steep blue stellar
continuum and strong Balmer absorptions (EW(H$\beta$) = $6.1$\AA\ and
EW(H$\alpha$) = $3.4$\AA\ (rest-frame)); these features are typical of
young post-starburst stellar populations with strong star-formation
activity rapidly quenched in the last Gyr \citep{couc+1987,pogg+1999}.

In addition to JO204, other sources are detected in the MUSE datacube; we
characterized all of them to assess any possible contamination in the
analysis of JO204. Some of these sources are marked in
Fig. \ref{fig:rgb}.  The object marked as S1 is a galaxy at
$z\simeq0.052$ with a typical post-starburst spectrum; the H$\beta$
equivalent width measured by integrating the spectrum on a circular
region of $1\arcsec$ is EW(H$\beta$)=$8.6$ \AA\ (rest-frame) in absorption.  The
spectrum of the diffuse object region around S2 shows no emission
lines and weak Balmer lines in absorption at a redshift similar to that
of S1; this is most likely a passive galaxy member of A957
cluster. S3 is a Milky Way star.  The spectra of these three sources
show no emission lines, and therefore they do not affect the analysis
of JO204 gas emission.  However, the absorption features of S1, S2 and
S3 could potentially affect the analysis of the stellar component of
JO204 and therefore we masked out the regions around these three
sources for the analysis of the JO204 stellar component (see
Sect. \ref{sec:stellarspec}).  S4 is a star forming galaxy at
$z\simeq0.36$. The strong [OIII]5007 emission line of this
galaxy is redshifted at a wavelength not far from the expected
location of H$\alpha$ emission of JO204.  We visually inspected the
MUSE datacube and found no signal of any possible emission from JO204
around S4; however, to avoid any possible confusion in the analysis,
we masked out a region of 5\arcsec around S4.  A few background
galaxies are also visible in Fig. \ref{fig:rgb}. These sources do not
significantly affect any feature in the spectrum of JO204 or its tail.

The observed datacube was corrected for Galactic extinction adopting a
value $E_{B-V}=0.037$ extracted from \cite{schl+2011} reddening maps
and the \cite{card+1989} extinction law.  In this paper we are mostly
interested in the low-surface brightness emission from the diffuse gas
in the tail; therefore, to increase the signal-to-noise ratio (SNR),
we applied a 5 pixel wide boxcar filter to the datacube; this means
that each spaxel (at each wavelength), was replaced with the average
value of the $5\times5$ neighboring spaxels.  This procedure does not
significantly alter the spatial resolution of the data, as the filter
size is similar to the seeing ($0\farcs8$, see Sect. \ref{sec:obs})
and it conserves the total flux.
A first estimate of the emission line fluxes and gas radial velocity
map was extracted from the resulting smoothed extinction-corrected
datacubes; to this purpose we fit a combination of 1D Gaussian
functions to the H$\beta$, [OIII]4959, [OIII]5007,
[OI]6300, H$\alpha$, [NII]6548, [NII]6583 and
[SII]6716, and [SII]6731 lines using the KUBEVIZ \citep{foss+2016}
software.

KUBEVIZ also output the continuum level for each line;
when no $H\alpha$ emission is detected, this value simply corresponds
to the mean value of the continuum in the $H\alpha$ spectral region (at the redshift of JO204).
To trace the distribution of the stellar component,
we extracted from the  $H\alpha$-continuum
map contour lines (in steps of 0.5 mag/arcsec$^2$).
The resulting isophotes are shown in Fig. \ref{fig:av}
and most of the maps in this paper\footnote{The
isophotes 
 also trace the continuum
emission from interlopers, such as S1, S2, S3, and S4 (see Fig. \ref{fig:rgb}).}.

The spectrum of the central region of JO204 in Fig. \ref{fig:spec1}
shows a blue wing in the [OIII] lines; a visual inspection of the
datacube confirms the presence of more than one gas component in the
central region ($\sim 5$kpc) of JO204, likely due to the AGN outflows.  Besides
the central regions, however, the line profiles are well described by a
single component and therefore we will base the analysis of this paper
on the simple single-component fit results.  The presence of multiple
gas component in the central regions of JO204 does not affect any of the
main conclusions of this paper, which is mainly focused on the
properties of the stripped gas. An analysis of the central
active region of JO204 and other GASP galaxies using a multi-component fit to
MUSE spectra is presented in \cite{pogg+agn}.

\subsection{Modeling the stellar component}\label{sec:stellarspec}

For the whole sample of galaxies in the GASP survey it is of primary
importance to disentangle the gas and the stellar components in the
observed spectra.  The stellar component is analyzed by masking out the
spectral regions of the principal emission lines.  As a first step, we
performed an adaptive spatial binning of the MUSE datacube using the
Voronoi tessellation method of \cite{capp+2003}, requiring a minimum
SNR of 10 on each binned region. As discussed in the previous section,
we masked out the regions around S1, S2, and S3 (see
Fig. \ref{fig:rgb}). The stellar radial velocity for each Voronoi bin
was then computed with the pPXF code \citep{capp+2004}. The resulting
stellar velocity map is presented and discussed in
Sect. \ref{sec:kin}.

The spatially resolved properties of the stellar populations were
obtained using the spectral fitting code SINOPSIS \citep{frit+2017} as
described in \cite{pogg+2017}.  This code combines different single
stellar populations spectra to reproduce the observed
equivalent widths of the main absorption and emission lines, as well as
the continuum in various wavelength bands. The outputs of SINOPSIS are
maps of: (i) stellar mass; (ii) average star formation rate (SFR) and total
mass formed in four age bins\footnote{Young: $<2\times10^7$yr, recent:
  $[2\times10^7$yr, $5.7\times10^8$yr], intermediate-age
  $[5.7\times10^8$yr $,5.7\times10^9$yr], and old: $>5.7\times 10^9$yr.}; (iii)
luminosity-weighted and mass-weighted stellar ages.  In addition, it produces a best-fit
model datacube for the stellar-only component.
The derivation of the SFR as a function of the cosmic time from optical spectra is both model-dependent
and prone to uncertainties, mainly due to the parameters' degeneracy, which is intrinsic to the problem. The ability
and the limitations of SINOPSIS in recovering the SFH have been extensively tested by 
\cite{fritz07,fritz11} using both mock and real data. The choice of the four age bins was done 
based both on the results of this comparison, and on the differences between the spectral features of 
simple stellar population spectra as a function of age. When applied to spatially resolved data, like these ones,
the method has proved to be even less prone to uncertainties, as highlighted in \cite{frit+2017}.

\subsection{Emission-only datacubes}\label{sec:emonly}

\begin{figure}[t]
 \includegraphics[width=\hsize]{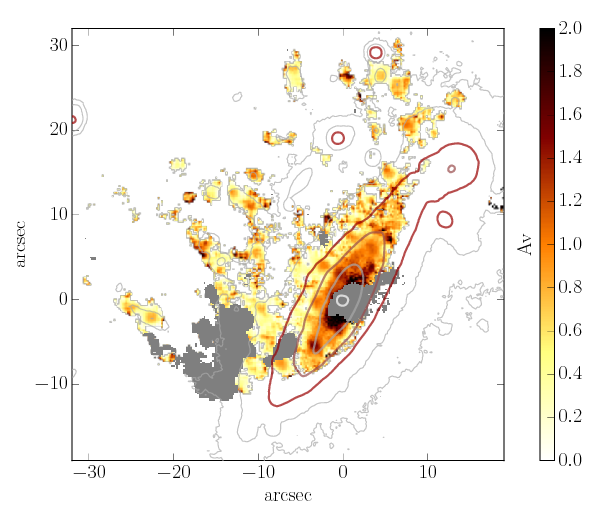}
  \caption{Extinction map by dust internal to JO204.  $A_V$ is
    estimated assuming that the gas is ionized by young stars,
    therefore the values obtained for regions ionized by shocks
    (AGN/LINERS) could not be reliable. These regions (see
    Sect. \ref{sec:ion}) are shaded in gray.
    Isophotes of
    H$\alpha$ continuum in steps of 0.5 mag/arcsec$^{-2}$ are shown to
    trace the stellar emission.
    \label{fig:av}}
\end{figure}

The spectra of the gas component were obtained by subtracting the
best-fit stellar model obtained with SINOPSIS to the
extinction-corrected datacube.  We run KUBEVIZ on the resulting
emission-only datacube to refine the gas radial velocity maps and
obtain the corrected emission fluxes for all emission lines.  
The measured fluxes were then corrected for extinction by dust internal to
JO204. The correction was carried out assuming an intrinsic 
$H\alpha/H\beta=2.86$ and the \cite{card+1989} extinction law.
Figure \ref{fig:av} shows that the highest values of the extinction in JO204
are found at the center ($A_V\gtrsim2$ mag); we found values
around $A_V\sim1.0$  mag on the galaxy disk, out to $\sim10\arcsec$.
The extraplanar dust in the gas tail has a clumpy distribution, with 
compact regions with $A_V\sim1.0$ mag; outside these dust-rich knots the extinction
is generally lower, with $A_V\lesssim0.5$ mag. In the following sections we will
show that the high-extinction regions
are bright compact sources of $H\alpha$ emission, as usually seen
for star-forming HII regions.

\begin{figure*}[t]
 \includegraphics[width=\hsize]{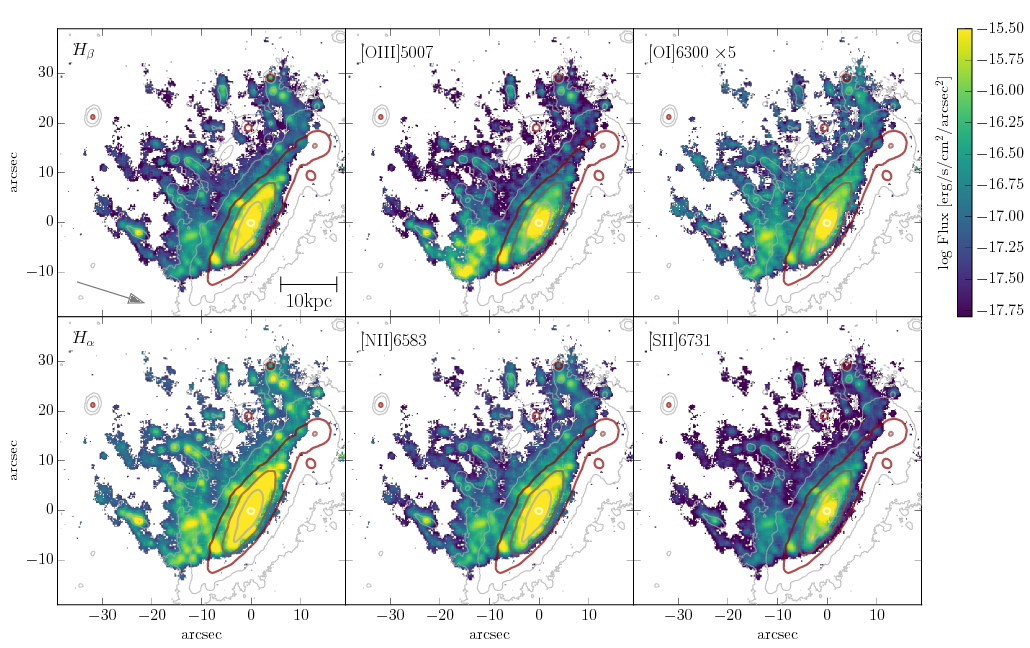}
  \caption{Absorption-corrected flux emission maps of the principal
    emission lines. The [OI] line is on average weaker than the
    others; to enhance the faint structures and keep the same scale
    for all maps, [OI] is displayed multiplied by a factor 5.  Only
    spaxels with SNR of H$\alpha$ flux $>5$ are shown.
    $H\alpha$ continuum isophotes are the same as in Fig. \ref{fig:av}.
    The arrow in the upper-left panel
    points to the cluster center.
    \label{fig:emmap}}
\end{figure*}

The maps of the dust-corrected emission-only fluxes for the principal
emission lines are shown in Fig. \ref{fig:emmap}, masking all spaxels
where the SNR of the H$\alpha$ flux is less than 5.
The flux maps
obtained for JO204 confirm the extraordinary sensitivity of MUSE,
which allows detection of faint sources, down to
$F_{\mathrm{H}\alpha}\simeq10^{-17.5}$ erg s$^{-1}$ cm$^{-2}$
arcsec$^{-2}$ with a SNR=5. This value is the typical detection limit
of the GASP survey \citep{pogg+2017} and is similar to the one obtained
by other MUSE observations \citep{fuma+2014,foss+2016}.

\begin{figure}[!ht]
 \includegraphics[width=\hsize]{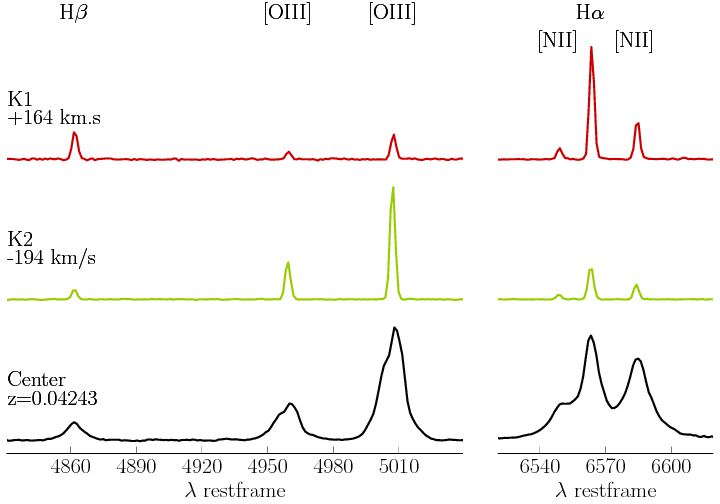}
  \caption{Pure 
emission spectra of the center of JO204 and of regions
    K1 and K2 (see Fig.\ref{fig:rgb}).  Spectra were obtained by
    subtracting the stellar component from the observed spectra. The
    spectra of the center of JO204 and K2 are typical of regions
    ionized by AGN activity (strong [OIII] and [NII]), while the K1
    emission is typical of star-forming clumps.
\label{fig:spec1}}
\end{figure}

The emission line maps in Fig. \ref{fig:emmap} clearly unveil the
bright tail of ionized gas in JO204.  This tail is developed along the
direction perpendicular to JO204 disk and is populated by a large
number of compact knots with strong Balmer line emission. 
The strong [OIII] emission region
around K2 (see Fig. \ref{fig:rgb}) is clearly visible in
Fig. \ref{fig:emmap}; the spectrum of K2 is shown together with those
of K1 and the JO204 center in Fig. \ref{fig:spec1}; the spectra in the central
region of the galaxy have emission lines with dispersion $\sim 300$km/s  and
[OIII]/H$\beta$ and [NII]/H$\alpha$ line ratios typical of AGN
emission; the line ratios in K2 spectrum have similar characteristics,
while K1 shows features typical of photoionized star-forming HII
regions.  The characterization of the ionizing mechanisms of the gas
component in JO204 is presented in Sect. \ref{sec:ion}.

The stellar disk, traced by the continuum isophotes in
Fig. \ref{fig:emmap}, extends at least up to $\sim 30$\arcsec\ from
the center. In the outer part of the disk there is no evidence of any
emission from the gas component, which is detected along the JO204 major
axis only in the inner 10$\arcsec$. 

\subsection{Gas and stellar kinematics}\label{sec:kin}

\begin{figure}[!ht]
\centering
\includegraphics[width=.95\hsize]{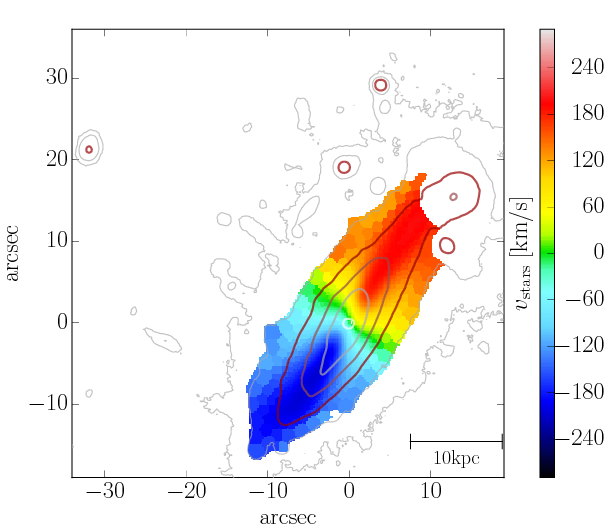}\\
\includegraphics[width=.95\hsize]{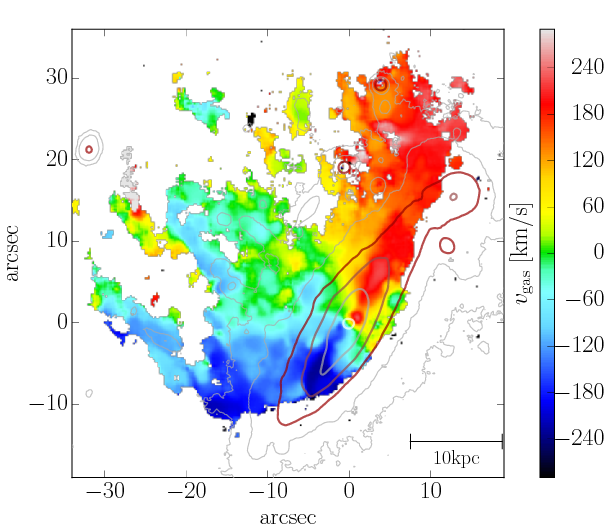}
\caption{Velocity map of JO204 stellar component {\it (upper panel)} and gas
  {\it (lower panel)}. H$\alpha$-continuum isophotes as in Fig. \ref{fig:av}.
  \label{fig:vmaps}}
\end{figure}

The velocity map of the gas and stellar components of JO204 is shown
in the upper panel of Fig. \ref{fig:vmaps}.  Radial velocities are
referred to the value obtained for the stellar component at the center
of the galaxy, which is $z=0.04243$.  The value of the stellar radial
velocity measured by pPXF in each Voronoi bin was smoothed using the
two-dimensional local regression techniques (LOESS) as implemented in
the Python code developed by
M. Cappellari\footnote{\url{http://www-astro.physics.ox.ac.uk/~mxc/software}}.

The radial velocity distribution of the stellar component of JO204 is
symmetric and it indicates a regular disk rotation.  The maximum
projected rotation velocity is $\sim\pm200$km/s out to
$\sim15$--20\arcsec\ (13--18kpc).  The regular rotation, together with
the regular morphology of the stellar isophotes, confirms that the
mechanism stripping the gas does not affect the stellar component, as
expected for ram-pressure stripping from the ICM.  The H$\alpha$
velocity map (lower panel of Fig. \ref{fig:vmaps}) shows that, on the
galaxy disk, the gas kinematics follows the stellar kinematics.  The most
notably difference between the gas and stellar velocity maps is the compact
region with $v_\mathrm{gas}\sim200$km/s located $\sim
1$\arcsec\ North-West of the center, which will be discussed in the
following paragraph.  The kinematics of the gas tail follows the
rotation of the stellar disk with only marginal local perturbations.
This is also observed in other jellyfish galaxies
\citep{fuma+2014,pogg+2017} and it is interpreted as a consequence of
the high speed of the galaxy in the ICM.

\begin{figure}[!ht]
\centering
\includegraphics[width=.95\hsize]{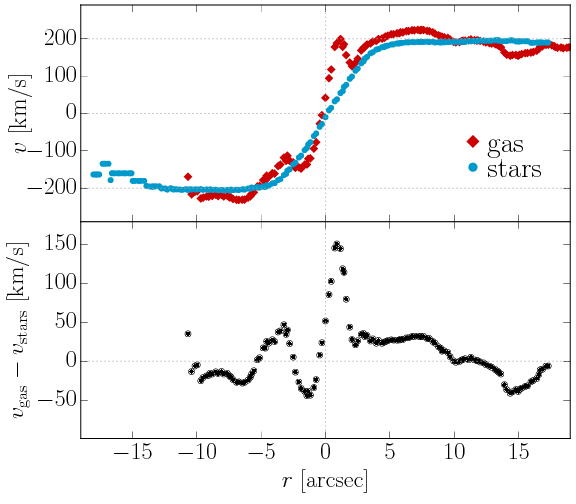}\\
  \caption{The {\it upper panel} shows the radial velocity for the gas
    and stellar component obtained along a slit aligned with the major
    axis of JO204.  The difference between the radial velocity of the
    two components is shown in the {\it lower panel}.
\label{fig:vslit}}
\end{figure}

To further analyze the kinematics of the two components, we extracted the
gas and stellar radial velocities along a narrow slit on the major axis
of JO204; the
position angle ($-30^\circ$) was estimated by fitting an ellipse to
the third brightest isophote in Fig. \ref{fig:emmap}.  The resulting
(projected) rotation curves are shown in Fig. \ref{fig:vslit}. 
The
gas component displays a double-hump rotation curve; the local minimum
and maximum are located at $\sim\pm1$\arcsec\ from the center of
JO204.  This could be the signature of a bar, or a
gas outflow from the AGN.  
The general shapes
of the two curves are similar, however, and show a  maximum
radial velocity of 200km/s; to test the robustness of these results
we performed a few tests, varying the angle of the slit used to 
trace  the gas and stellar radial velocity curves.
The stellar and gas
kinematics do not show significant variations in the inner ($\sim
10\arcsec$) disk other than the two aforementioned humps.
Lastly, we note that the symmetry of
the radial velocity curves shows that the rotation center (of both the stellar and the gas components)
corresponds to the 
center of the galaxy, as defined by the emission peak.

\subsection{Gas ionization mechanism}\label{sec:ion}

\begin{figure*}[!ht]
 \includegraphics[width=0.49\hsize,trim={.5cm 0 0 0},clip]{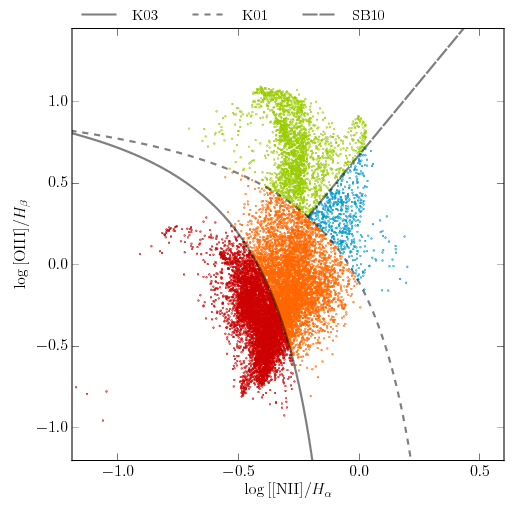}\hfill
 \includegraphics[width=0.49\hsize,trim={.5cm 0 0 0},clip]{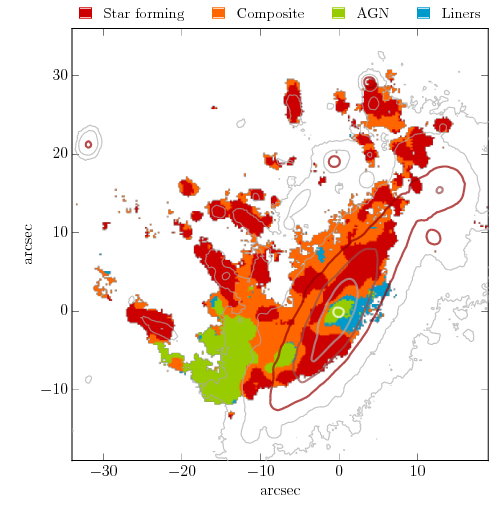}\\[1em]
 \includegraphics[width=0.49\hsize,trim={.5cm 0 0 0},clip]{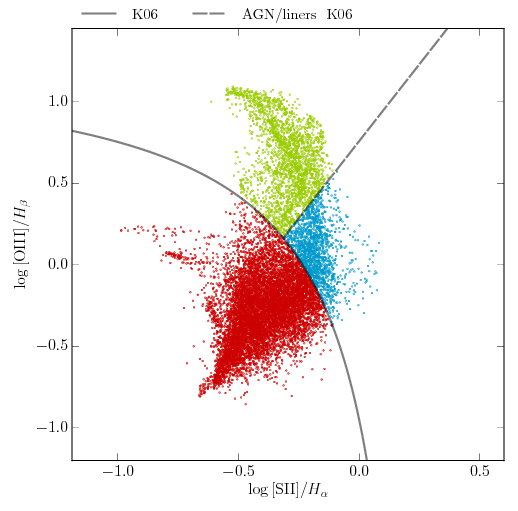}\hfill
 \includegraphics[width=0.49\hsize,trim={.5cm 0 0 0},clip]{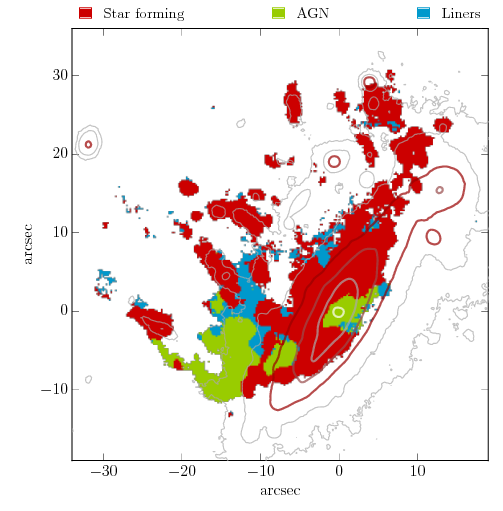}
  \caption{{\it Left panels}: BPT diagnostic diagrams for
[OIII]5007/H$\beta$ vs [NII]6583/H$\alpha$ ({\it top}) and 
[OIII]5007/H$\beta$ vs [SII]6717,6731/H$\alpha$ ({\it bottom}) line-ratios for all spaxels with fluxes measured with a SNR $>3$.
The lines used to separate star-forming (red dots), composite (orange), LINERs (blue) and AGNs (green) are from
\citet[][K03]{kauf+2003},
\citet[][K01]{kewl+2001},
\citet[][K06]{kewl+2006}, and
\citet[][SB10]{shar+2010}.
{\it Right panels}: corresponding maps showing the ionization mechanism as derived from the BTP diagrams.
The contours are stellar isophotes, as in Fig. \ref{fig:av}.
    \label{fig:bpt}}
\end{figure*}

To investigate the physical mechanism responsible for the gas
ionization we used the line-ratio diagnostic
Baldwin-Phillips-Terlevich \cite[BPT,][]{bpt1981} diagrams. Such
diagrams are commonly used to distinguish sources photoionized by hot
young stars from those ionized by LINERS or AGN.  We built
two BPT diagrams using [OIII]5007/H$\beta$ vs [SII]6717+6731/H$\alpha$\footnote{[SII]6717+6731 is the sum of the fluxes of the two [SII] lines.}
and [OIII]5007/H$\beta$ vs [NII]6583/H$\alpha$ line ratios obtained
for each MUSE spaxel with SNR$>3$.  Results are shown in Fig. \ref{fig:bpt}.  
Classification of star-forming/composite regions, LINERS and AGN was
performed following the results of
\citet{kauf+2003},\citet{kewl+2001,kewl+2006}, and \citet{shar+2010}.
The center of JO204 is dominated by the AGN emission.  Interestingly, there is also an extended
AGN-powered region 15-20 kpc away from the center; this 
is the  bright [OIII] region surrounding K2 in
Fig. \ref{fig:rgb} and it is likely an AGN ionization cone. The other
side of the cone could be associated with the region that extends
$\sim5$kpc from the center on the opposite direction and that is
another region that is classified as a LINER or AGN-powered in the two
BPT diagrams.

In addition to the center and the region being affected by the AGN-ionization cone, 
the gas in the disk and in the stripped tail is ionized by young stars.
As noted by \cite{pogg+2017}, these are very likely stars formed $10^7$ yr ago at most,
within the stripped gas. Support for this hypothesis is given in Sect. \ref{sec:blob} and \ref{sec:sfh}
where we show that the JO204 tail hosts many HII regions forming massive young stars
that can ionize the gas.

\subsection{Metallicity}\label{sec:met}

\begin{figure*}[!ht]
 \includegraphics[width=0.49\hsize]{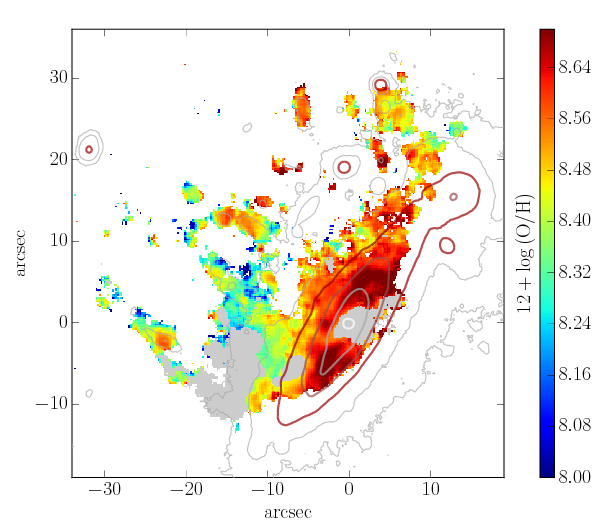}\hfill
 \includegraphics[width=0.49\hsize]{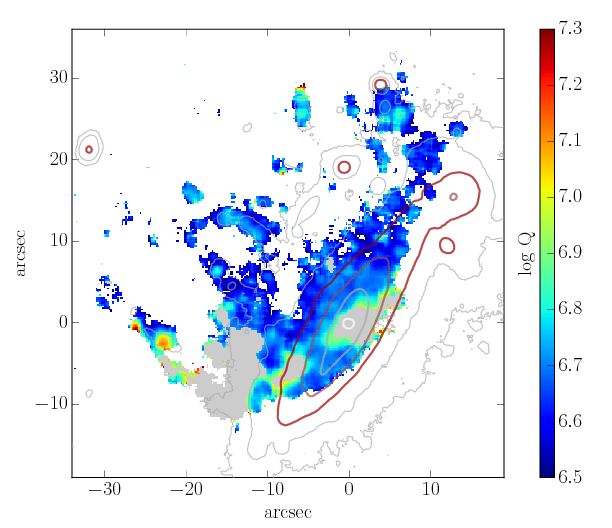}\\
  \caption{Metallicity ({\it left panel}) and ionization parameter ({\it right panel}) map.
    Regions ionized by LINERS/AGN according to the BPT diagram have been masked ({\it gray area}).
    Contours are stellar isophotes as in Fig. \ref{fig:av}.
    \label{fig:met}}
\end{figure*}

The gas metallicity and ionization parameter for each spaxel were
calculated using the {\it pyqz} Python
module\footnote{\url{http://fpavogt.github.io/pyqz/}} \citep{dopi+2013}
v0.8.2; the $\log(Q)$ and $12+\log(O/H)$ values are obtained by
interpolating from a finite set of diagnostic line ratio grids
computed with the MAPPINGS code.  We used a modified version of the
code (F. Vogt, priv comm.)  to implement the MAPPING IV grids that are
calibrated in the range $7.39<12+\log(O/H)<9.39$, which is broader
than the metallicity range covered by MAPPING V grids. 
Our results are
based on the [NII]6585 / [SII]6717+6731 vs [OIII]/[SII]6717+6731 \cite[see][]{pogg+2017}. The calibration applies only to star-forming
regions, therefore we masked out all regions classified as AGNs or
LINERs.  The resulting metallicity and ionization parameter maps are
shown in Fig. \ref{fig:met}.  The metallicity of the ionized gas
varies by $\sim 0.7$ dex.  The inner galaxy disk is more metal-rich
than the outskirts and the stripped tail, as
expected. Interestingly, the western side of the gas tail (at
coordinates $X\sim5^\prime$, $Y\sim 10^\prime$ in Fig. \ref{fig:met})
seems to be more metal-rich than the eastern side, as if it were stripped
from the inner --and hence more metal-rich-- regions of the galaxy disk. This
would in turn be explained if the galaxy velocity vector were 
not exactly perpendicular to the galaxy disk, but it would be pointing
South (in the direction of the negative y-axis in Fig. \ref{fig:met}).

This is the second paper based on the GASP survey in which we
analyze the physical properties of the stripped gas in a jellyfish galaxy;
the previous one presented data for JO206 \cite{pogg+2017}. Future papers
will compare the properties of GASP galaxies and of the ionized gas.
We note, however, that with respect to JO206, 
the ionized gas in JO204 is generally more
metal-rich; instead, both galaxies have very low ionization parameter
with $\log q<7$, as shown in the right panel of Fig. \ref{fig:met}.

\subsection{$H\alpha$ Knots}\label{sec:blob}

\begin{figure*}[!ht]
 \includegraphics[width=0.49\hsize]{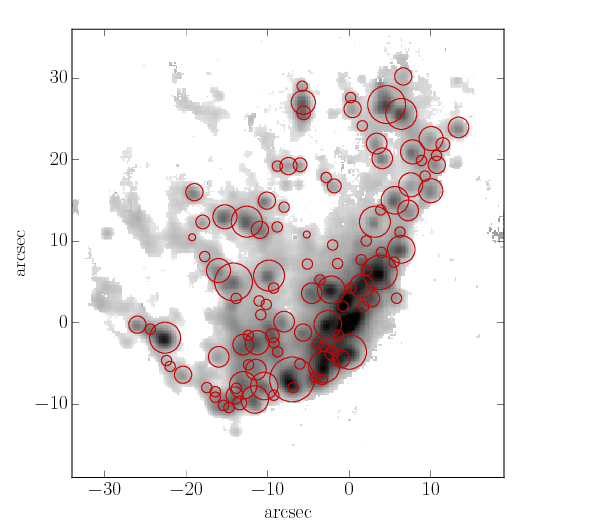}\hfill
 \includegraphics[width=0.49\hsize]{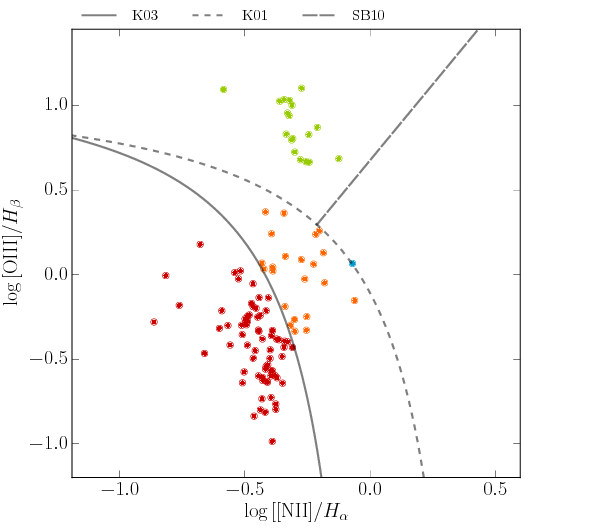}\\
 \includegraphics[width=0.49\hsize]{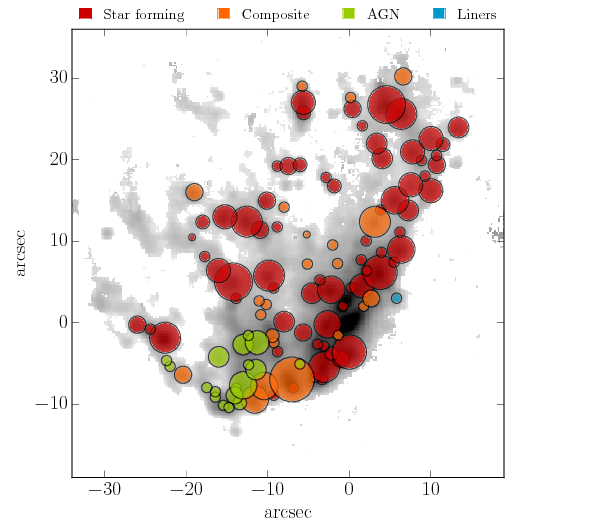}\hfill
 \includegraphics[width=0.49\hsize]{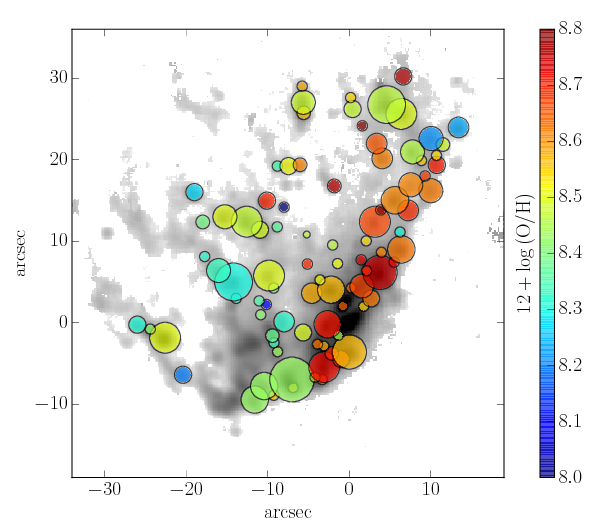}
  \caption{
  The upper-left panel shows the  111 $H\alpha$ knots detected in JO204.
    The location of each knot in the  BPT diagram is shown in the upper-right panel
    (the color-coding and lines are the same as in Fig. \ref{fig:bpt}) and it is 
       used to derive the ionization mechanism; results are
       shown in the lower-left panel.
   The metallicity of the  92 star-forming regions is shown in the lower-right panel.
           \label{fig:blob}}
\end{figure*}

The ionized gas in JO204 is very clumpy, and as in most GASP galaxies, it
presents many bright star-forming knots in the galaxy disk in and the
tail; using the procedure described in \cite{pogg+2017}, we identified and
measured the size and H$\alpha$ flux of 
111 individual H$\alpha$ knots in JO204; our measurements
provide reliable flux measurements, taking care of deblending the
contributions from overlapping sources.  KUBEVIZ was then run to
measure the integrated emission fluxes of all knots.  The nature of
these blobs was determined on the basis of BPT diagrams, as those
described in Sect. \ref{sec:ion}.  Figure \ref{fig:blob} shows that 92
of these knots are star-forming/composite regions.  Excluding the
central emission peak due to the AGN itself, all AGN-powered knots are
found in the eastern region, which was already discussed in the previous
sections.  The presence of a large number of star-forming HII regions
provides support to our conclusion, presented in the previous section,
that most of the gas in JO204 is ionized by hot stars.  HII regions
formed in situ in the gas tail seem to be a common feature of most
galaxies undergoing gas stripping
\citep{yagi+2013,foss+2016,pogg+2017}; however, this contrasts with the
case of NGC~4569, a jellyfish galaxy in the Virgo cluster:
\cite{bose+2016} found no star forming HII region in it and therefore
they speculate that the gas is mainly excited by mechanisms other than
photoionization (shocks in the turbulent gas, magnetohydrodynamic waves, and heat
conduction).

The lower-right panel in Fig. \ref{fig:blob} shows that the
metallicity of the blobs follows the metallicity distribution of the
diffuse gas.  The HII regions on the west side of the gas tail (up to
$X\sim 10$, $Y\sim15$) are more metal-rich ($12+\log(O/H)>8.6$ than
those on the east side; this could be due to the fact that these HII
regions formed from gas stripped from the center of JO204 rather than
from the outer disk; this in turn provides some indications that the
ram-pressure stripping is acting along a south-north direction, as
already noted in Sect. \ref{sec:met}. This would implicate that the
vector of the motion of JO204 across the ICM points downward on
our maps (~South).

\begin{figure*}[!ht]
 \includegraphics[width=\hsize]{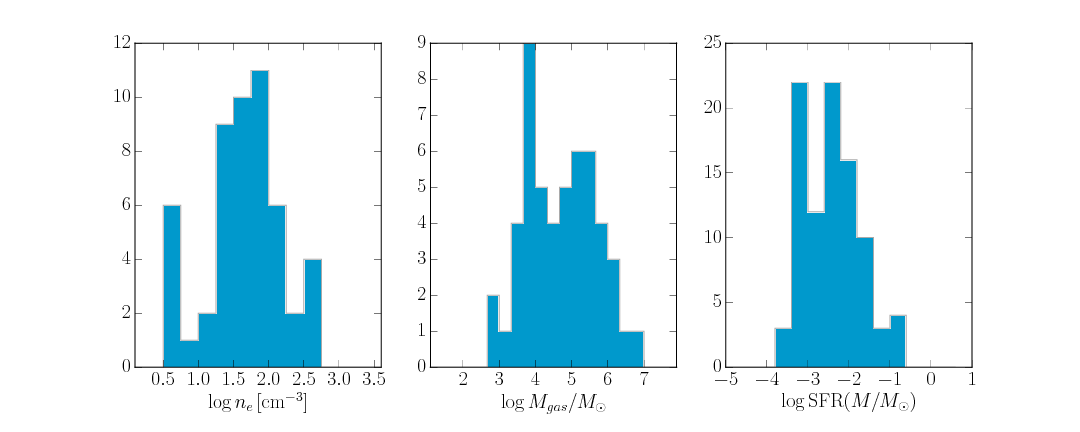}
  \caption{Distributions of gas density, mass and SFR of individual H$\alpha$ knots.
    \label{fig:blobmass}}
\end{figure*}

The ongoing SFR of each blob was calculated from the
H$\alpha$ absorption- and dust-corrected flux as in in \cite{pogg+2017}, adopting Kennicutt's
relation for a Chabrier IMF:
\begin{equation}
SFR=4.6\times10^{-42}L_{\mathrm{H}\alpha}
\end{equation}
The total SFR for the 92 star-forming regions is  $1.4\, M_\sun/yr$,
and the distribution of SFRs for all blobs is shown in Fig. \ref{fig:blobmass}.

To compute the gas mass of each star forming region,
we estimated the electron density
from the \cite{prox+2014} relation; this is based on the
$R=$[SII]6717/[SII]6737 line ratio 
and is valid in the range
$0.4<R<1.435$. We refer to \cite{pogg+2017} for a detailed
description of the procedure.
51 knots have  [SII] line ratios within the range where the density calibration applies.
Their mass distribution 
is shown in
Fig. \ref{fig:blobmass}.  The values we found range from $10^3$ to $10^7 M_\sun$
and the  median value is $5.3\times 10^4 M_\sun$.
The total gas mass of the 51 blobs is $1.8\times10^7M_\sun$.
This is one order of magnitude lower than the total gas mass of the star-forming blobs
in JO206 \citep{pogg+2017}. 
These results will be compared to those that are being obtained from other GASP galaxies
to assess the general properties 
of star forming regions in the stripped gas.

\subsection{Star Formation History}\label{sec:sfh}

An estimate of the total ongoing SFR can be obtained from the dust-
and absorption-corrected $H\alpha$ luminosity in the star-forming
regions; toward this end, we integrated the $H\alpha$ emission maps
masking out all AGN- and LINER-powered
regions (see Sect. \ref{sec:ion}) and found SFR=1.7 $M_\sun$/yr.  This
is a hard lower limit to the total SFR because the AGN region accounts
for about $45\%$ of the total $H\alpha$ luminosity. Our results,
however, provide reliable estimates of the order of magnitude of the
total SFR; using the results of the previous section, we can also
conclude that --excluding AGN-powered regions-- most of the
star-formation activity is taking place inside the compact HII regions.  We
also computed the total SFR outside the regions defined by the second,
third and the fourth brightest stellar isophotes that are shown in all
maps (e.g. Fig. \ref{fig:emmap}) and we found values of 1.5, 0.8 and 0.6
$M_\sun$/yr, respectively; the SFR outside the main body of JO204 is therefore
$\lesssim 0.5 M_\sun$/yr, which is much lower that the value found for
JO206 \citep[$\sim$ 1$-$2.5 $M_\sun$/yr,][]{pogg+2017}; despite the
spectacular tail of stripped gas and the large number of star-forming
regions in the tail, at least 2/3 of the star formation activity in JO204 is taking place
in the disk.

To estimate the total disk stellar mass, we integrated the spectra in the regions inside the three 
previously defined isophotes and ran SINOPSIS on the resulting spectra: 
the total mass inside the inner isophote is $3 \times10^{10} M_\sun$,
inside the third isophote we computed a total mass of $4 \times10^{10} M_\sun$ and found that the 
value inside the outer isophote is the same, within the uncertainties that are typically $\sim10$--$20\%$,
showing that the stellar mass outside the outer isophote is negligible.

\begin{figure*}[!ht]
 \includegraphics[width=\hsize]{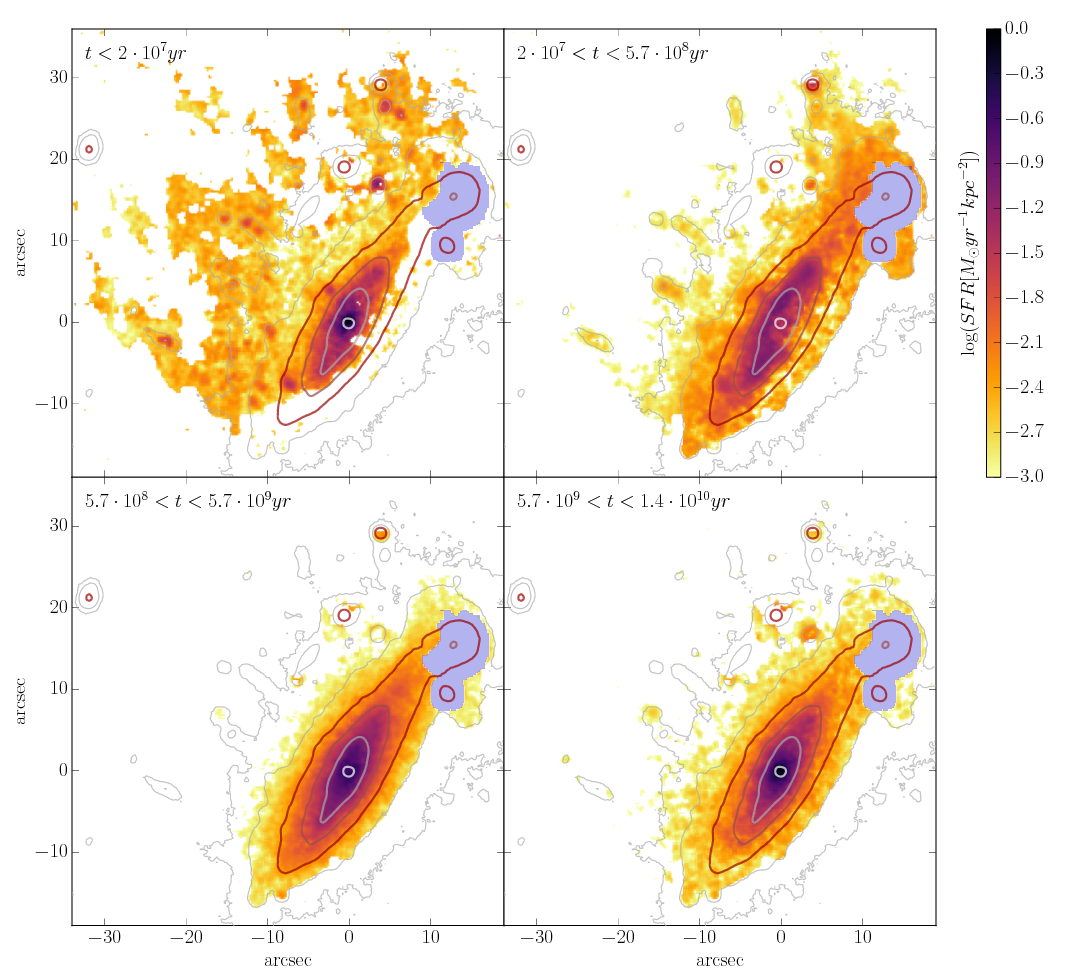}
  \caption{Maps of the average star formation rate per kpc$^2$ in four different age bins.
    The areas shaded in blue are regions contaminated by interloper galaxies;
    these were
    excluded from the SINOPSIS analysis
    because we could not define the redshift of either the gas or the stellar component.
    The contours show the H$\alpha$-continuum just as in Fig. \ref{fig:av}.
    \label{fig:sfh}}
\end{figure*}

The spatially resolved star formation history of JO204 was
reconstructed using the output of SINOPSIS.  The results are shown in
Fig. \ref{fig:sfh}.  The ongoing star formation rate, as traced by
stars born in the last $2\times 10^7$yr, is extremely intense in the
inner disk, particularly in the very central region of JO204, as
already discussed at the beginning of this section.  We detected no
ongoing star-formation in the outer disk, where the gas has been
already completely stripped and the star formation activity has been
totally quenched. Some moderate ongoing star formation is detected
everywhere across the stripped gas tail; the distribution is clumpy
and the star-formation activity is concentrated in the knots
identified in the previous section.

The recent star formation (between 20 and 570 Myr, the second youngest age bin) 
is not distributed like the ongoing star formation ($<20$ Myr).
It is rather intense in the outer disk and less peaked toward the center of the galaxy with respect
to the ongoing star formation. Figure \ref{fig:sfh} suggests a modest presence of stars born in this age bin
in the regions of the tail closer to the main body of the galaxy. Star formation activity 
in the external tail of JO204 is negligible between 20 and 570 Myr (Fig. \ref{fig:sfh}, upper left panel).

The distribution of the stellar populations older than 500 Myr is regular (see the two bottom panels in Fig. \ref{fig:sfh}, as expected for
undisturbed disk galaxies. No old stellar population is detected in
the tails of JO204.

The SFH in the outer disk is peaked within the second age bin; this is
expected, as we have already shown that the spectra in these regions have
post-starburst features, indicating vigorous star formation activity
that was abruptly interrupted less than 1 Gyr ago.  The oldest
stars in the (inner) tails were born within the last $\sim 500$ Myr; this should 
therefore be the epoch of onset of the gas-stripping and the beginning of the
gas tail. 
The burst of star formation in the outer disk should have 
therefore been induced by the compression due to ram-pressure in this
region.

\section{JO204 environment}\label{sec:env}
JO204 is a member of A957, a relatively low-mass cluster with an X-ray
luminosity $L_X=7.8\times10^{43}$ erg/s in the 0.1-2.4 keV band
\citep{ebel+1996}.  From the analysis of OmegaWINGS spectroscopy,
\cite{more+2017} measured a total cluster mass $M=4.4 \times 10^{14}
M_\sun$, a velocity dispersion $\sigma_v = 640$ km/s, a systemic
cluster redshift $z= 0.0451$, and a size $R_{200} = 1.5$ Mpc.  For
JO204 we measured a systemic redshift $z=0.04243$, therefore the
galaxy moves with a radial velocity of $-800$km/s with respect to A957,
which corresponds to 1.25 times the cluster velocity dispersion.
JO204 is located in the central region of the cluster, at a projected
clustercentric distance of $2\farcm1$, which corresponds to 132 kpc or
$0.08 R_{200}$.

\begin{figure*}[t]
 \includegraphics[width=\hsize]{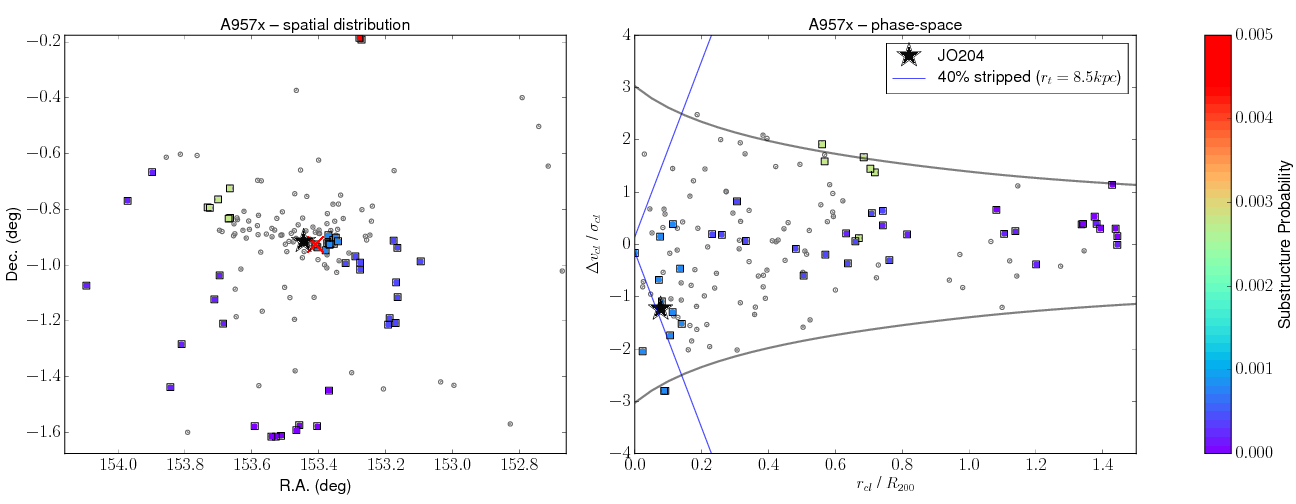}
  \caption{
  The position in the sky of the spectroscopic members of A957 from OmegaWINGS (small gray circles), JO204 (black star), and the BCG (red cross). Squares correspond to identified substructures, which have been color-coded according to their probability of being random fluctuations (i.e. values close to zero indicate highly significant substructure detections; Biviano et al. in preparation). Right: position vs. velocity phase-space diagram with symbols as in the left panel. The gray curves show the escape velocity in a \cite{nfw} halo. The dashed lines intersecting with the position of JO204 correspond to 30\% of the total gas mass stripped in JO204 by the ICM in a Virgo-like cluster (see the text for details). The solid blue line corresponds to the expected stripping from the extent of the (truncated) $H\alpha$ emission (64\% of the total gas mass stripped). 
  \label{fig:psd}}
\end{figure*}

The left panel of Figure \ref{fig:psd} shows the location of JO204 within the cluster (star).  
Cluster members of A957 plotted as gray circles for reference, along with four galaxy groups
identified by a substructure analysis (colored squares, Biviano et al. in preparation).
Although several substructures are present in the cluster, JO204  is not associated with any of them.
The right panel of Figure \ref{fig:psd} further shows the location of JO204 in a projected clustercentric distance vs.
velocity phase-space diagram. In these kind of plots, gravitationally bound galaxies congregate inside a
trumpet-shaped region enclosed by the escape velocity (gray curves), with the majority of galaxies in the
low $r$ and low $|\Delta_v|$ region.  The bulk of galaxies near the
center were accreted a long time ago and thus can be thought of as
virialized. The most recent members of the clusters, on the other hand,
can be distributed across the whole diagram, and tend to have higher
absolute velocities (and velocity dispersion) than the virialized population
\citep[see][]{hain+2015,jaff+2015,jaff+2016}. 

Using the (projected) phase-space position of JO204, we can estimate the expected
ram-pressure intensity exerted by the cluster  
as
\begin{equation}\label{eq:pram}
P_{ram} = \rho_{ICM} \times \Delta v_{cl}^{2}
\label{eq:Pram}
\end{equation}
\citep{gunn+1972},
where $\rho_{ICM}(r_{cl})$ is the radial density profile of the ICM,
$r_{cl}$ is the clustercentric distance, and $\Delta v_{cl}$ is the velocity of
the galaxy with respect to the cluster. 
Since A957 is a low-mass cluster similar to the well-studied Virgo cluster,
we adopt the ICM density values of Virgo from  \cite{voll+2001} and
obtain a value for the ram-pressure of $P_{ram} = 4.2 \times10^{-13} N m^{-2}$.  We note that this is a lower limit to $P_{ram}$, as it is based on the projected distance and velocity of the galaxy with respect to the cluster. In particular, considering the extent of the galaxy tails in the sky, the projected velocity is likely an  underestimation of the full velocity.

To test how much ram-pressure stripping is caused on
the disk of JO204, we compute the anchoring force of the galaxy assuming an
exponential disk density profile for the stars and the gas components
($\Sigma_{g}$ and $\Sigma_{s}$ respectively) as:

\begin{equation}
\Sigma=  \left(\frac{M_{d}}{2 \pi r_d^2}\right)  e^{-r/r_d},
\label{sigma0}
\end{equation}

where $M_{d}$ is the disk mass, $r_d$ is the disk scale-length and $r$ is the
radial distance from the center of the galaxy.
For the stellar component of JO204 we adopted a disk mass
$M_{d,\star}=4\times10^{10}M_{\odot}$,
and a disk scale-length $r_{d,\star} = 4.2$~kpc, obtained by fitting
the light profile of the galaxy. 
For the gas component we assumed a total mass $M_{d,gas} = 0.15 \times
M_{d,\star}$ \citep{popping}, 
and scale-length $r_{d,gas} = 1.7 \times r_{d,\star}$ \citep{bose+2006}.

The above values can be used to compute the  anchoring force in the disk at different $r$ following
\begin{equation}
\Pi_{gal} = 2 \pi G \Sigma_{g} \Sigma_{s}
\label{eq:Pi}
\end{equation}

In the phase-space diagram of Fig. \ref{fig:psd} we plot several stripping lines, corresponding to
different stripping (or truncation) radii, $r_t$.
In each case, the fraction of remaining gas mass can be computed as  
\begin{equation}
f=1+\left[e^{-r_t/r_d}\left(\frac{-r_t}{r_d} -1\right)\right] \label{eq:frac}
\end{equation}

We find that the condition for stripping ($P_{ram}>\Pi_{gal}$) is met at
a truncation radius $\sim 8.5$~kpc (40\% of the total gas mass stripped). This radius coincides remarkably well with the extent of the H$\alpha$ emission with respect to the stellar disk.

Overall, our analysis suggests that the gas in JO204 is being significantly stripped by a ram-pressure force  perpendicular to the disk.

\section{Hydrodynamical $N$-body simulations}\label{sec:sims}
To model JO204's evolution, we used the simulation code \textit{GIZMO} \citep{gizmo} in its ``meshless finite-mass'' implementation of the hydrodynamics-solver.

\begin{figure}
    \includegraphics[width=\hsize]{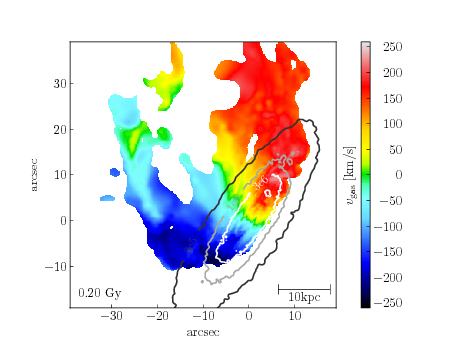}
    \caption{Velocity map of the closest-match simulation in our grid. Superimposed are contours of equal stellar mass (per pixel).}
    \label{fig:simvrot}
\end{figure}

We followed \cite{rammerge} in modeling a constant ram-pressure in a wind-tunnel-like setup, acting on galaxy initial conditions generated to resemble the pre-interaction properties of JO204. 
The galaxy initial conditions generator is described in \cite{mdg}.
For JO204, these initial conditions are comprised of dark matter particles distributed in accordance with a Hernquist profile\footnote{\cite{hernprof} profile: $ \rho_{\mathrm{dm}}(r) = \frac{M_{\mathrm{dm}}}{2\pi} \frac{a}{r(r+a)^3}$ with $a= r_s\,\sqrt{2[\mathrm{ln}(1+c)-c/(1+c)}$ and $c=r_{200}/r_s$ where $r_s$ is the scale length of a regular NFW halo \citep{nfw}.}.
Concentrically embedded in this dark matter halo are exponential disks, one each for the stellar and gaseous components.
These galaxy models are then pre-aged in isolation for one Gyr in order for star formation to self-regulate and the disks to stabilize before they can be placed in the wind-tunnel. 

\begin{table}
    \centering
    \caption{ %
        Parameters of the closest-matched simulation in the ensemble.
        First are the galaxy initial condition parameters, followed by the properties of the simulated ICM, and lastly, the viewing angle. Note that some parameter (in particular the disk scale length) are subject to moderate change during pre-aging and simulation in the windtunnel.
    }
    \label{tab:simparams} \vspace{0.5em}
    \begin{tabular}{lcrl}
        \hline
        Parameter & Symbol & Value & unit \\
        \hline
        halo concentr. & $c$              & 12               & - \\
        halo circ. vel.& $v_{200}$        & 120              & km/s \\
        spin parameter & $\lambda$        & 9.5              & \% \\
        stellar mass   & $M_{d,\star}$    & $4\cdot10^{10}$  & M$_\odot$ \\
        gas mass       & $M_\mathrm{d,gas}$ & $6\cdot10^9$   & M$_\odot$ \\
        disc scalelen. & $r_\mathrm{d}$   & $4.2$            & kpc \\
        disk height    & $z_0$            & 0.2              & $r_\mathrm{d}$ \\
        \hline
        ICM density & $\rho_\mathrm{ICM}$ & $5\cdot10^{-28}$ & g~cm$^{-3}$ \\
        ICM vel.       & $|\Delta v_\mathrm{ICM}|$ & 1000             & km/s \\
        ICM incl.      & $\beta$          & 30               & deg \\
        viewing angle  & $\alpha$         & -18              & deg \\
        \hline
    \end{tabular}
\end{table}

The initial gas fraction $f_\mathrm{gas}$ of our models was estimated from the stellar mass using the relations in \cite{popping}, as in Section~\ref{sec:env}.
We varied the remaining parameters of the halo and those of the ICM in a grid of simulations.
The radial velocity of JO204 with respect to A957 served as a lower limit for the ICM velocity relative to the model galaxy ($|\Delta v_\mathrm{ICM}| > |\Delta v_\mathrm{cl}|$).
Similarly, the clustercentric distance in the viewing plane was assumed as a lower limit for the three-dimensional distance of JO204 to A957.
The latter was used to obtain limits for the simulated density of the ICM ($\rho_\mathrm{ICM}$).
We assumed a beta-profile\footnote{Beta-profile: $\rho_\mathrm{icm}(r)\approx\rho_0 [ 1 + \left( r/r_c \right)^2 ]^{-3/2\beta}$, \citep{beta}} 
and used the parameters in \cite{ebel+1996} and \cite{ku} for A957, giving us a rather large range of $\sim 2.2\cdot10^{-28}$ to $\sim 1.4\cdot10^{-27}$~g/cm$^3$ for $\rho_\mathrm{ICM}$.
The initial parameters of the best-match simulation and the fit parameters of the stellar disk are listed in Tab. \ref{tab:simparams}.

Figure \ref{fig:simvrot} shows the velocity field of our closest-matched simulation.
To generate Fig. \ref{fig:simvrot}, we interpolated the LOS-velocity of the simulated gas particles with a nearest-neighbor-based smoothing-kernel and integrated the resulting 3D map along the LOS (similar to \citealt{stei+2016}). 
Pixels with a (LOS-integrated) gaseous mass below a threshold of $\sim 30$~M$_\odot$ were removed.
These pixels stem from sparse regions, where the particle density in the simulation is low. These pixels would otherwise lead to artifacts and inaccurate results at the outskirts of the gaseous wake in Fig. \ref{fig:simvrot}. 

The stripping fraction was estimated by engulfing the disk stars in a cylinder of a thickness equal to twice the disk height.
The cylinder is centered on the disk stars. 
The stripping fraction is then taken as the mass ratio of the gas outside this cylinder vs. total gas mass: $$f_\mathrm{strip} = \frac{M_\mathrm{gas,out}}{M_\mathrm{gas,cyl}} \sim40\%$$
This is in remarkable agreement with the fraction of stripped gas derived from analytical considerations (Sect. \ref{sec:env}).

There is an overall agreement between the gas velocity map in Fig.\ref{fig:simvrot}
of the simulated galaxy and the velocity map of the JO204 gas component in Fig.\ref{fig:vmaps};
the hydrodynamic simulations also confirm the reliability of our conclusion 
that the onset epoch of ram-pressure stripping is a few $10^8$ yr ago.

\section{Summary and conclusions}\label{sec:conclusions}

In this paper we presented MUSE observations from the ESO Large
Programme GASP, a survey of candidate gas stripping galaxies.
 This
paper is focused on JO204, a member of the relatively low-mass cluster A957.
The total exposure time of our MUSE observations is 4050 s; the
detection limit on the final reduced datacube is
$F_{\mathrm{H}\alpha}\simeq10^{-17.5}$ erg s$^{-1}$ cm$^{-2}$
arcsec$^{-2}$ (SNR=5).

We summarize the main results of this paper as follows.
\begin{itemize}
\item
MUSE data unveiled a prominent tail of ionized gas stripped from
the main body of the galaxy; 
it has an extension of $\sim30$kpc in the direction
perpendicular to the major axis of JO204 and it is $\sim45$kpc wide.
\item
The spectra of the outer disk show no emission by ionized gas and
strong Balmer-line absorption. These features are
typical of post-starburst stellar populations, whose star formation activity has been rapidly quenched at some point $\lesssim 1$ Gyr ago.
\item
The kinematics of the stellar component does not reveal any significant
sign of perturbation from a regular disk rotation.
\item
The kinematics of the gas in the disk follows the rotation of the
stellar component; 
this rotation
is also retained --with minor local perturbation-- by the gas and the HII regions in the tail.
\item
Apart from the AGN and its ionization cone, most of the gas is ionized
by in situ star formation.
\item
We identified 92 star forming blobs in JO204, which account for most of the total $\sim2M_\sun/yr$ ongoing star formation rate.
\item
The ongoing SFR outside the main body of the galaxy is estimated to be $0.6 M_\sun/yr$.
\item
The star formation activity in the tail began during the last 500 Myr;
in the external regions of the tail we did not detect stars
older than $2\times10^7$ yr.
\end{itemize}

The interpretation of these data was supported by an analysis of
the phase-space diagram and basic ram-pressure stripping prescriptions \citep{gunn+1972};
we also found a further support to the conclusions drawn from MUSE by comparing the observed gas kinematics with the results from a set of hydrodynamical simulations.
We conclude that JO204 is at the first infall in A957, in a highly radial orbit;
a few $10^8$ yr ago, ram-pressure started stripping gas.
The consequent compression of the gas induced a burst of star formation
in the disk; the gas was then rapidly and completely removed from the outer disk, quenching the star formation. 
Ram pressure stripping proceeded outside-in and it is  still active;
we estimate that $40\%$ of JO204's total gas mass has been stripped so far. 
The gas removed from the galaxy's main body collapsed in star-forming regions
throughout the gas tail; excluding the AGN-powered regions,
we found that about a third of the ongoing SFR (excluding the AGN-powered regions)
is taking place outside the main galaxy body; this is very similar to
the fraction found in JO206 by \cite{pogg+2017}.

This galaxy is at a similar stage of stripping as another of the spectacular jellyfish galaxies of GASP, JO201 \citep{bell+2017}, but with the stripping being observed in the plane of the sky rather than along the line of sight.

MUSE observations is providing valuable data for probing the ionized gas
and the stellar populations of GASP galaxies, as shown in this paper;
these observations triggered a number of follow-up programs to probe
all the gas phases and to study the connection between gas stripping and
the star formation activity. JO204 was selected as a target for GASP follow-up programs
aimed at studying the molecular gas (APEX, Moretti et al. in prep), atomic gas (JVLA),
and X-ray gas (Chandra) in this galaxy.

\acknowledgments
We would like to thank the anonymous referee for helping improve the paper.
This work is based on observations collected at the European Organisation for Astronomical Research in the Southern Hemisphere under ESO programme 196.B-0578. 
This work made use of the KUBEVIZ software, which is publicly available at \url{http://www.mpe.mpg.de/?dwilman/kubeviz}. 
We warmly thank Matteo Fossati and Dave Wilman for their invaluable help with KUBEVIZ, and Frederick Vogt for useful discussions and help with optimizing pyqz. 
We acknowledge financial support from PRIN-INAF 2014. 
J.F. aknowledges financial support from UNAM-DGAPA-PAPIIT IA104015 grant, Mexico.
This work was co-funded under the Marie Curie Actions of the European Commission (FP7-COFUND). 
B.V. acknowledges the support from 
an  Australian Research Council Discovery Early Career Researcher Award (PD0028506).
\bibliographystyle{aasjournal}
\bibliography{gasp_JO204I}



\facilities{VLT (MUSE), VST (OmegaCAM), AAT (AAOmega).}

\software{KUBEVIZ, ESOREX, SINOPSIS, CLOUDY, pyqz, IDL, Python.}

\end{document}